\documentclass[fleqn,twoside]{article}

\usepackage{natbib}
\usepackage{espcrc2}

\usepackage{graphicx}

\usepackage{amssymb}

\begin{document}



\title{X-ray Emission Diagnostics from the M87 Jet}

\author{E. S. Perlman\address[UMBC]{Joint Center for Astrophysics, University of
Maryland, Baltimore County}
A. S. Wilson\address[UMCP]{Astronomy Department, University of
Maryland, College Park}}

\begin{abstract}

We use Chandra, HST and VLA observations of M87 to investigate the physics of
X-ray emission from AGN jets.  We find that X-ray hotspots in the M87 jet occur
primarily in regions with hard optical-to-X-ray spectra and lower than average 
polarization. Particle injection appears to be required both continuously in
the jet sheath as well as locally at X-ray hotspots.

\end{abstract}

\maketitle


\section{Introduction and Observations}

The M87 jet is the nearest (16 Mpc, giving a scale of $1'' = 78$ pc) and
highest surface-brightness jet in the optical, radio and X-rays.  As such it
makes an excellent prototype for studying jet physics.  The jet shows only
modest differences in morphology between the optical and radio: in the optical
the jet appears knottier and is more concentrated along the centerline (Sparks,
Biretta \& Macchetto 1996).  Baade (1956) showed that its radio-optical
emission was synchrotron radiation, on the basis of its high polarization. 
X-ray emission from the jet of  M87 was first cleanly separated from Virgo
cluster X-ray emission by  {\it Einstein} (Biretta, Stern \& Harris 1991), but
until the launch of {\it Chandra}, there was essentially no information on its
X-ray morphology.  

Deep Chandra observations (Wilson \& Yang 2002) of M87 were taken 2000 July
29-30 with ACIS-S.   We compare those data to  HST observations (Perlman et al.
2001) taken 1998 February and April, which include 7 bands between 0.3-2.05
$\mu$m wavelength, and to HST (V band) and radio (15 GHz) polarimetry 
observations (Perlman et al. 1999), which were obtained in 1995 May and 1994
February respectively.  We convolved the HST and VLA images with Gaussians to a
common resolution of $0.5''$ (FWHM) for morphology comparisons; however, we
have chosen to leave the polarimetry data at full ($0.2''$) resolution to 
bring out relevant details. 

We show in Figure 1 the image of the jet in the 0.3-1.5 keV band (where the PSF
is smallest in size and varies the least), after maximum-entropy deconvolution
using a monochromatic 1 keV PSF. Deconvolution of the {\it Chandra} data
improved the resolution from $0.84''$ to $0.54''$ (FWHM), enabling us to
resolve knot HST-1 from the nucleus and better separate other features.
Superposed as contours on this are (at left) an HST I-band image of the jet,
Gaussian smoothed to $0.5''$ resolution, and (at right) an unsmoothed HST
polarization image ($0.2''$ resolution).  In Figure 2 we show the run of X-ray
flux, along with softness ratios SR1= F(0.3-1 keV)/F(1-3 keV) and SR2=F(1-3
keV)/F(3-10 keV) as well as $\alpha_{ox}$, the optical-to X-ray spectral index.

A full accounting of our work will appear as Perlman \& Wilson (2003).  We
refer the reader to that paper for details on our data reduction and
deconvolution procedures.  Here  we summarize some of the findings,
particularly as respects the issue of particle acceleration.

\begin{figure}

\centerline{\includegraphics*[angle=90,scale=0.55]{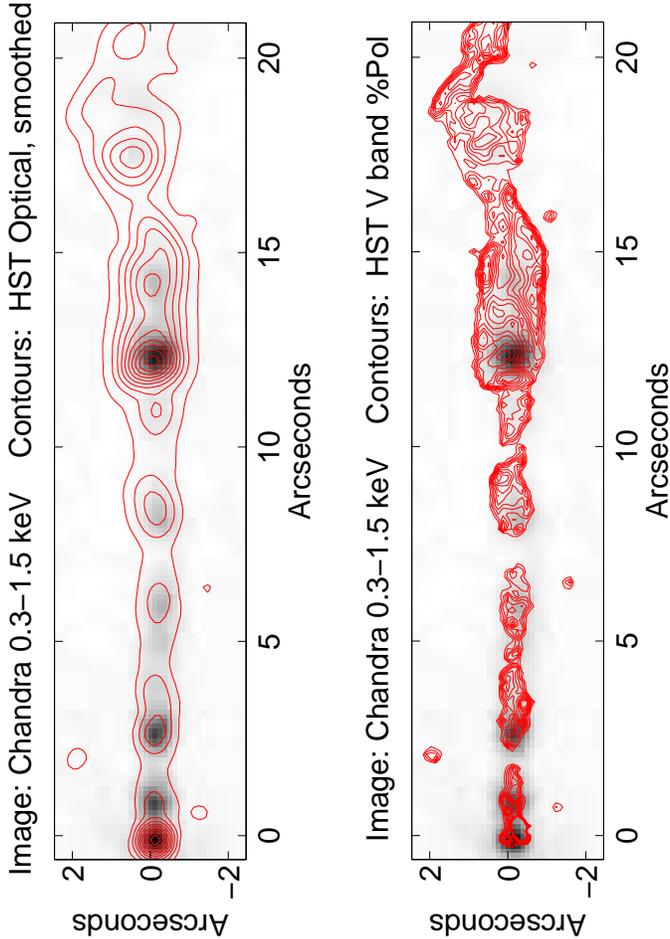}}

\caption{{\it Left panel.} Deconvolved {\it Chandra} 0.3--1.5 keV  image of the 
jet with contours from a smoothed {\it HST} I-band (F814W) optical image.  In
both cases, the scaling is by the square root of image values.  {\it Right
panel.} Deconvolved {\it Chandra} 0.3--1.5 keV  image of the 
jet with contours from a full-resolution HST optical polarimetry image.  The
greyscaling is identical to the above; however, the contours begin at 5\%
polarization and go up to 60\%.}

\end{figure}

\section{Jet Morphology, Spectrum and Polarimetry}

The optical and X-ray emission of the jet (Figure 1) track fairly
closely in most regions; however, some significant differences are seen.  In
particular, two bright X-ray hotspots are located where there is no
corresponding optical hotspot: in the interknot region between knots D and E,
and in the interknot region between knots C and G (both were mentioned in
Marshall et al. 2002 and Wilson \& Yang 2002).  In addition, the X-ray maxima
of a two knots (E and F) are located a few tenths of an arcsecond upstream of
their optical maxima. 

All components are consistent with an X-ray spectral index $\alpha_x=1.3$. 
This has been reported before for other components by Wilson \& Yang and
Marshall et al.; however,  our analysis is the first to isolate knot HST-1 (see
also later) as well as an attempt to fit a spectral index map (not shown). 
There do, however, appear to be variations in the softness ratios (Figure 2),
which decrease significantly at distances $<4''$ from the nucleus.  The
variations in SR1 are consistent with a gradually increasing column in the
innermost 300 pc of the jet.  Significant improvements in the spectral fits are
obtained with columns $N(H) = 8.7 \times 10^{20} {\rm cm^{-2}}$ for the nucleus
(previously noted  by Wilson \& Yang) and $6.5 \times 10^{20} {\rm cm ^{-2}}$
for HST-1.  The variations in SR2, however, cannot be explained by an
increasing column at these moderate $N(H)$ values.  Rather, they suggest a
hardening of the jet spectrum at high energies, but there are too few photons
in the hard band to adequately constrain this.  

There is considerably more variation in $\alpha_{ox}$ than in $\alpha_x$.  Knot
HST-1 has a much harder optical-to-X-ray spectrum than any other knot, with
$\alpha_{ox}=0.9$. The remainder of the inner jet has $\alpha_{ox}=1.2-1.5$, with
steeper spectra at larger distances from the nucleus.  In addition to this
pattern, one also sees small flattenings at the positions of the X-ray
hotspots. The additional flattenings are suggestive of localized particle
acceleration at the knot maxima (see \S 3).  The jet's optical-to-X-ray
spectrum becomes even steeper beyond knot A ($12.4''$ from the nucleus),
reaching  $\alpha_{ox}=1.8$ in knot C.  Interestingly, the knot A-B-C complex
does knot show significant spectral hardening near the knot maxima.

\begin{figure}

\centerline{\includegraphics*[scale=0.5]{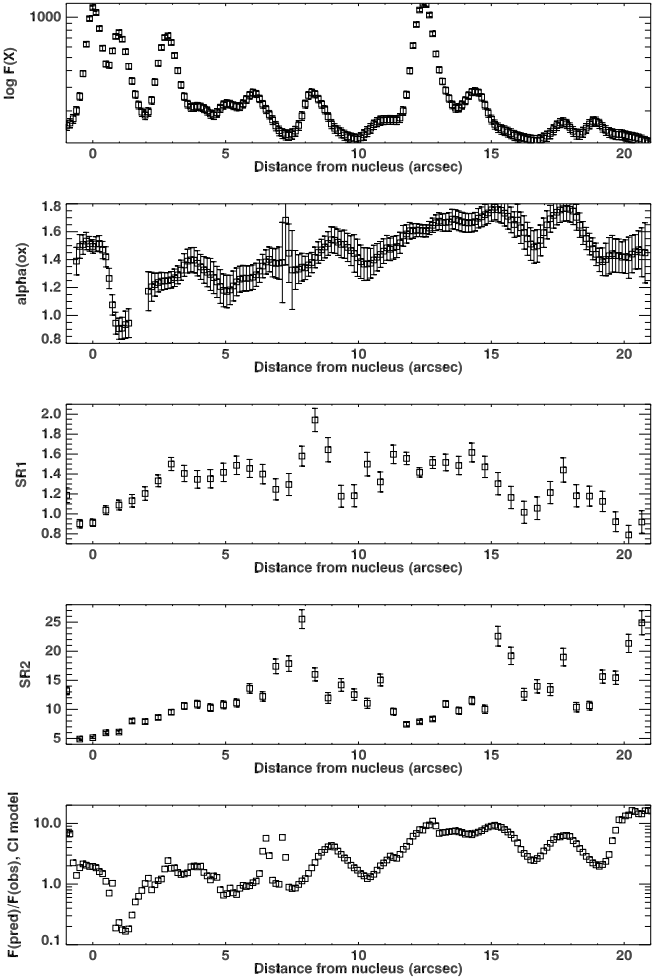}}

\caption{{\it Top Panel.} The run of X-ray flux in the deconvolved {\it
Chandra} image. Plotted against this (with the same distance scale) are:  {\it
Second from top.} Optical-to-X-ray spectral index, {\it Third and Fourth from
the top.} Softness ratios SR1 and SR2, and {\it Bottom.} The comparison of
predicted to observed 1 keV flux from the continuous injection synchrotron
model.}

\end{figure}

X-ray flux is strongly anti-correlated with optical polarization (Figure 1),
although the details of the relationship differ in the inner and outer jet.  In
all the knots in the inner jet, the X-ray flux peak is located at a local
polarization minimum with $P<20\% $, similar to the optical flux-optical
polarization anti-correlation noted in Perlman et al. (1999).  Immediately
upstream from inner jet knot maxima we see increases in polarization and
magnetic fields to the jet, while downstream from the knot maxima we see
increased polarization but magnetic fields parallel to the jet.  In the outer
jet, the anti-correlation between X-ray flux and optical polarization is
weaker.  Knot A's maximum is located at a local minimum in polarization
although unlike the inner jet knots the polarization there is still appreciable
(35\%) rather than consistent with zero.  The peak of knot B is also located in
relatively low polarization regions, but there are also polarization minima in
the knot A-B-C complex  which do not correspond to X-ray maxima. In addition,
the X-ray peak of knot C is located in a fairly high polarization region.  

\section{Physical Implications}

X-ray synchrotron emitting particles have lifetimes of only a few to tens of
years assuming near-equipartition magnetic fields (e.g., Meisenheimer, R\"oser
\& Schl\"otelburg 1996, Heinz \& Begelman 1997), meaning that in situ particle
acceleration is required to produce the observed X-ray emission extending over
a jet 7000 ly long.  Can we identify loci of particle acceleration?  In Figure
2 one notices an excellent correlation (at least in the inner jet) between the
loci of X-ray flux maxima and the loci of flat optical-to-X-ray spectrum
regions.  Such spectral changes are suggestive of local particle acceleration
in the knot maxima.  However our modeling shows that to be insufficient to
explain the observed X-ray emission.

We fit the radio to optical data with synchrotron spectrum models, and then use
the models to predict the X-ray flux and spectral index in each pixel. We have
done this using the code of Leahy (1991) and Carilli et al. (1991).  Three
models were fit: (1) the Jaffe \& Perola (1972) model, which assumes no
continuous particle injection and but includes pitch-angle reisotropization;
(2) the Kardashev (1962) and Pacholczyk (1970) model, which assumes neither
particle injection nor pitch-angle scattering; and (3) a continuous injection
(CI: Heavens \& Meisenheimer 1987) model, under which a power law distribution
of electrons is continuously injected.  We show in the bottom panel of Figure 2
the ratio $F_{pred}/F_{obs}$ at 1 keV for the continuous injection model.  The
Jaffe \& Perola model is not shown because it underpredicts the X-ray flux by
orders of magnitude at many places and predicts an exponential decay of the
X-ray spectral index (not observed), while the Kardashev-Pacholczyk model is
not shown because it consistently underpredicts the X-ray flux by large factors
and predicts too steep a spectral index.

Two main patterns can be seen in this plot.  In most of the jet (except for
knot HST-1), $F_{pred}/F_{obs} \sim 1-10$, meaning that particle acceleration
occurs within 100-10\% of the volume of the jet. There is a gradual increase in
$F_{pred}/F_{obs}$ as the distance from the nucleus increases.  Small increases
in  $F_{pred}/F_{obs}$ occur at knot maxima in the inner jet.  This suggests
that if particle acceleration occurs within the knot maxima, as suggested
by the  optical-to-X-ray (this paper) and optical spectra (Perlman et al. 2001)
at these points, the loci of particle acceleration must be much smaller than 
{\it Chandra} can resolve.

Knot HST-1 appears to be a special case.  It is the only place in the jet where
the continuous injection model significantly {\it underpredicts} the X-ray flux.
The reason for this is unclear.  Knot HST-1 is known to be very active, with 
superluminally moving components (Biretta et al. 1999).  Recently, HST-1 has
shown blazar-like X-ray 'flaring' (Harris, these proceedings).  It is possible
that this variability plays a part in the anomalous $F_{pred}/F_{obs}$ seen in 
HST-1.  Alternately, an extra emission component could be present in the jet
at higher energies; however, our modeling does not have sufficient angular
resolution to test this hypothesis.

Importantly, we do {\it not} see large departures in the value of
$F_{pred}/F_{obs}$ in inter-knot regions.  Moreover, the spectral models that
do not include particle injection or acceleration still underpredict the X-ray
emission at these loci by large factors. Thus even in inter-knot regions we
find evidence of continuous particle injection, most likely operating in the
sheath of the M87 jet.  A similar conclusion was reached by Jester et al.
(2001) for 3C 273 on the basis of radio-optical data. 

Interestingly, we see different polarization signatures in the regions where
our modeling indicates {\it in situ} particle acceleration.  As shown in Figure
1 and Perlman et al. (1999), the sheath of the M87 jet exhibits high
polarization and magnetic fields parallel to the jet, while the knot maxima 
have either low or no polarization.  Thus while the knots appear to be
shock-like features (Perlman et al. 1999), where Fermi acceleration may be the
dominant process,  a different mechanism may operate in the sheath.

\end{document}